# Large-scale digital phenotyping: identifying depression and anxiety indicators in a general UK population with over 10,000 participants

Yuezhou Zhang[1,*], PhD; Callum Stewart[1], PhD; Yatharth Ranjan[1], MSc; Pauline Conde[1], BSc; Heet Sankesara[1], BSc; Zulqarnain Rashid[1], PhD; Shaoxiong Sun[1,2], PhD; Richard J B Dobson[1,3,4,5,#], PhD; Amos A Folarin[1,3,4,5,#,*], PhD

[1]Department of Biostatistics & Health Informatics, Institute of Psychiatry, Psychology and Neuroscience, King's College London, UK
[2]Department of Computer Science, University of Sheffield, Sheffield, UK
[3]Institute of Health Informatics, University College London, UK
[4]NIHR Maudsley Biomedical Research Centre, South London and Maudsley NHS Foundation Trust, London, United Kingdom
[5]NIHR Biomedical Research Centre at University College London Hospitals, NHS Foundation Trust, London, United Kingdom

[#] These authors contributed equally
[*] Corresponding authors

Emails of corresponding authors: yuezhou.zhang@kcl.ac.uk (Yuezhou Zhang) and amos.folarin@kcl.ac.uk (Amos A Folarin).

Postal address of corresponding authors: Department of Biostatistics & Health Informatics, Institute of Psychiatry, Psychology and Neuroscience, King's College London, London, SE5 8AF, United Kingdom

**Abstract**

**Background:** Digital phenotyping offers a novel and cost-efficient approach for managing depression and anxiety. Previous studies, often limited to small-to-medium or specific populations, may lack generalizability.

**Methods:** We conducted a cross-sectional analysis of data from 10,129 participants recruited from a UK-based general population between June 2020 and August 2022. Participants shared wearable (Fitbit) data and self-reported questionnaires on depression (PHQ-8), anxiety (GAD-7), and mood via a study app. We first examined the correlations between PHQ-8/GAD-7 scores and wearable-derived features, demographics, health data, and mood assessments. Subsequently, unsupervised clustering was used to identify behavioural patterns associated with depression or anxiety. Finally, we employed separate XGBoost models to predict depression and anxiety and compared the results using different subsets of features.

**Results:** We observed significant associations between the severity of depression and anxiety with several factors, including mood, age, gender, BMI, sleep patterns, physical activity, and heart rate. Clustering analysis revealed that participants simultaneously exhibiting lower physical activity levels and higher heart rates reported more severe symptoms. Prediction models incorporating all types of variables achieved the best performance ($R^2$=0.41, MAE=3.42 for depression; $R^2$=0.31, MAE=3.50 for anxiety) compared to those using subsets of variables.

**Limitations:** Data collection during the COVID-19 pandemic may introduce biases, despite controlling for pandemic-related variables.

**Conclusion:** This study identified potential indicators for depression and anxiety, highlighting the utility of digital phenotyping and machine learning technologies for rapid screening of mental disorders in general populations. These findings provide robust real-world insights for future healthcare applications.

# 1. Introduction

Depression and anxiety are the most prevalent mental health disorders worldwide and leading causes of disability (Collaborators, 2022). These mental disorders are associated with adverse outcomes such as premature mortality, reduced quality of life, impaired occupational function, and increased suicide risk, thereby imposing significant burdens on society and individuals (Liu et al., 2020; Santomauro et al., 2021). Traditional detection and monitoring methods, such as interviews and questionnaires, are often susceptible to subjective recall bias and fail to capture day-to-day fluctuations in behaviours and mental health status, which can delay effective treatment (Ben-Zeev and Young, 2010). Moreover, there is a stark shortage of mental health professionals—approximately nine psychiatrists per 100,000 people in high-income countries and merely 0.1 per million in low-income regions (Murray et al., 2012; Oladeji and Gureje, 2016). The recent global COVID-19 pandemic has further highlighted the critical importance of remote healthcare, particularly when access to traditional health services is limited (Bufano et al., 2023; Santomauro et al., 2021). Consequently, there is a growing need for innovative, remote, and effective methods to monitor and manage depression and anxiety (Torous et al., 2020).

Ubiquitous digital devices such as smartphones and wearables provide a cost-effective means to gather data on individual behaviours, physiological signals, and mood states, providing insights into daily human experiences (Mohr et al., 2017; Torous et al., 2016). Digital phenotyping has emerged as an innovative approach that leverages data from these digital devices to characterise behaviours and health conditions (Insel, 2017; Mohr et al., 2020). This non-invasive, remote, and highly scalable approach has driven numerous mobile health (mHealth) studies for mental health, especially during the COVID-19 pandemic (Anikwe et al., 2022; Huckins et al., 2020; Ueafuea et al., 2020; Zhong et al., 2023). Focusing specifically on depression and anxiety, these studies collected real-world data to explore indicators related to these prevalent disorders, with some leveraging advanced machine learning algorithms to predict their severity (Abd-Alrazaq et al., 2023; De Angel et al., 2022; Rohani et al., 2018). Notably, passive measurements of behavioural and physiological patterns, including sleep routines, physical activity levels, and heart rate variability, have shown significant associations with depression and anxiety (Chalmers et al., 2014; De Angel et al., 2022; Moshe et al., 2021; Okobi et al., 2023). This evolving field holds the potential to transform mental health management by integrating continuous, objective data analysis into preventive and therapeutic strategies.

Despite these advancements, these studies had several limitations, including the use of small to medium datasets (De Angel et al., 2022) and homogeneous or specific populations (e.g., college students (Wang et al., 2014; Xu et al., 2023)). This limits the generalizability of their findings and may be the reason for some inconsistent results reported in different cohorts (Currey and Torous, 2022; De Angel et al., 2022; Rohani et al., 2018). Furthermore, the performance of machine learning models varied across datasets, and their broader applicability remains uncertain (Abd-Alrazaq et al., 2023; De Angel et al., 2022). Given that the prevalence of depression and anxiety varies across demographics (Akhtar-Danesh and Landeen, 2007; Bayram and Bilgel, 2008) and is influenced by factors such as physical comorbidities (Cimpean and Drake, 2011) and mood fluctuations (van Rijsbergen et al., 2013), there is a critical need for more comprehensive investigations of the indicators related to these conditions and testing the feasibility of machine learning prediction models in large-scale general populations.

The present study used a cross-sectional design and analysed data from a large-scale mHealth dataset encompassing over 10,000 participants from a general UK-based population (Stewart et al., 2021). This study aimed to (1) identify indicators of depression and anxiety through wearable-derived features, demographic data, health metrics like body mass index (BMI), and mood assessments, and (2) evaluate the feasibility of using these features to predict the severity of depression and anxiety through machine learning techniques and interpret the impact of different modal features on prediction outcomes.

## 2. Materials and methods

### 2.1 Dataset

This work analysed data from a large-scale observational mHealth study, Covid Collab, in which 17,667 participants self-enrolled between June 2020 and August 2022 via the study smartphone app (Stewart et al., 2021). While the app was available globally, the majority of participants (95%) were recruited from within the UK (Stewart et al., 2024). Participants shared their Fitbit wearable data through the RADAR-base, an open-source mHealth platform (Ranjan et al., 2019). They were also encouraged to regularly complete questionnaires related to mental health, mood, and COVID-19 symptoms via smartphone. Details of the study procedures and data collection are described in the study's protocol (Stewart et al., 2021). The Covid Collab study received ethical approval from the King's College London's PNM Research Ethics Panel (LRS-18/19-8662), with participants providing electronic consent through the study app.

### 2.2 Measures

2.2.1 Depression and anxiety assessments

Participants' depression and anxiety symptoms were assessed biweekly using the smartphone-conducted eight-item Patient Health Questionnaire (PHQ-8 (Kroenke et al., 2009)) and seven-item Generalized Anxiety Disorder (GAD-7 (Spitzer et al., 2006)). The total scores on these assessments could range up to 24 and 21 respectively, with higher scores indicating more severe symptoms. PHQ-8 and GAD-7 are commonly used self-reported questionnaires for rapid screening depression and anxiety (Kroenke et al., 2009; Spitzer et al., 2006).

2.2.2 Baseline variables: demographic and health factors

This study considered various demographic variables including age, ethnicity, employment status, marital status, and living situation. Health variables included BMI, physical comorbidities, and recent COVID-19 infections (diagnosed over the past week, month, and three months). These variables were collectively referred to as "baseline variables" throughout the rest of this paper.

2.2.3 Mood variables

Participants' mood was assessed twice weekly through a simple task within the study app, where participants rated their daily Arousal (from -1 for tiredness to 1 for wakefulness) and Valence (from -1 for displeasure to 1 for pleasure) using a sliding button. Mood variables closest to the date prior to each PHQ-8 and GAD-7 assessment were selected due to varying participation rates.

2.2.4 Fitbit features

Fitbit data was processed in two steps: daily feature extraction and biweekly feature aggregation. Daily features included sleep metrics (e.g., duration, onset, and offset times), activity metrics (e.g., caloric consumption, durations of active and sedentary periods), step metrics (e.g., total steps, and statistics of hourly steps), and heart rate metrics (e.g., hourly statistics and heart rate during activity). In addition to extracting features over the entire day, we also divided each day into four time periods and extracted features for each of them: night (00:00-06:00), morning (06:00-12:00), afternoon (12:00-18:00), and evening (18:00-24:00). To link participants' behavioural and physiological characteristics with depression and anxiety assessments, we calculated the mean and standard deviation of daily features over the 14 days preceding each PHQ-8/GAD-7 assessment. In total, 46 aggregated biweekly features were extracted (detailed in Supplementary Table 1). The supplementary methods section provides a detailed description of the feature extraction process.

**2.3 Data inclusion criteria**

Based on our previous research, to ensure the quality of features, we only extracted daily features if the day's missing rate was less than 20%, and calculated aggregated biweekly features if at least eight days of valid daily data were available within the 14 days preceding each assessment (Sun et al., 2023; Zhang et al., 2024b). The inclusion criteria for PHQ-8 and GAD-7 assessments in this study were as follows: (1) completeness, i.e., all sub-questions were answered; (2) at least one type of aggregated biweekly feature (sleep, activity, step, or heart rate) was extracted; and (3) for participants with multiple assessments, we selected the first that met the previous two criteria for the cross-sectional study design.

**2.4 Association analysis**

Associations between the extracted variables and PHQ-8/GAD-7 scores were examined using the Kruskal-Wallis test (Ostertagova et al., 2014) for categorical variables, and the Spearman correlation coefficient (Spearman, 1961) for continuous variables. The Benjamini-Hochberg method (Benjamini and Hochberg, 1995) was used for multiple comparison correction.

**2.5 Clustering analysis**

We conducted an unsupervised clustering analysis on the extracted biweekly Fitbit features to explore different behavioural and physiological patterns among participants and their links with depression and anxiety. In contrast to bivariate association analysis, clustering enables the examination of relationships between multiple feature combinations and mental health outcomes. Initially, mean imputation and standardization were conducted for data preprocessing. Subsequently, principal component analysis (PCA) was performed for dimensionality reduction, which transformed the original features into principal components (PCs) representing linear combinations of these features (Granato et al., 2018). PCs were selected for clustering analysis until the cumulative explained variance exceeded 80%. To achieve robust and stable clustering results, K-means clustering (Syakur et al., 2018) was executed 500 times with selected PCs, using the elbow method to determine the optimal number of clusters (k). This approach utilized different centroid seeds and selected clustering results with the lowest within-cluster sum of squares for the final output. After clustering, PHQ-8 and GAD-7 scores were compared across the identified clusters to discern patterns associated with the severity of depression and anxiety.

**2.6 Prediction models**

We aimed to predict the severity of depression and anxiety for "new participants" (those not included during the training phase). The XGBoost model was chosen for its robustness in handling non-linear relationships and missing data (Chen and Guestrin, 2016). We developed separate prediction models for depression and anxiety, using the PHQ-8 or GAD-7 scores as the outcomes respectively, with extracted features as predictors. Note, due to the high correlation between depression and anxiety, and them both being derived from questionnaire data, we avoided using GAD-7 to predict PHQ-8 and vice versa. The data were split into a 70% training set and a 30% test set, stratified by age, gender, and PHQ-8/GAD-7 scores to ensure similar distributions between the sets. Parameter tuning was performed via 5-fold cross-validation on the training set, and predictions were made on the unseen test data. The coefficient of determination ($R^2$) and mean absolute error (MAE) were used for model evaluation. Additionally, the SHapley Additive exPlanations (SHAP) method (Lundberg and Lee, 2017) was utilized to interpret feature importance and their impact on the predictions for PHQ-8 and GAD-7 scores. We tested model performance across all features and then conducted an ablation study to assess the predictive power of each category of features (baseline variables, mood variables, and Fitbit features). To minimize random bias, all models were repeated 50 times with different random seeds.

## 3. Results

**3.1 Data overview**

Our analysis included 10,129 participants selected based on the data inclusion criteria. The selected cohort had a median (IQR) age of 48.0 years (36.0-58.0), with 7,008 (69.3%) participants being female. The median PHQ-8 score was 6.0 (2.0-10.0), and the median GAD-7 score was 4.0 (1.0-8.0). Detailed demographic information is available in Table 1.

**3.2 Associations of depression and anxiety with extracted variables**

Significant differences in depression and anxiety scores were observed across various baseline variables. PHQ-8 and GAD-7 scores decreased with increasing age, with the 18-30 age group reporting median (IQR) scores of 9.0 (5.0–14.0) for PHQ-8 and 7.0 (4.0–13.0) for GAD-7, and the 70+ age group reporting scores of 3.0 (1.0–6.0) and 1.0 (0.0–4.0) for PHQ-8 and GAD-7 (Figure 1a; $p < 0.001$). Gender differences revealed that female participants exhibited higher median PHQ-8 and GAD-7 scores of 6.0 (3.0–11.0) and 5.0 (2.0–9.0) respectively, than males (PHQ-8: 4.0 [2.0–8.0]; GAD-7: 3.0 [0.0–6.0]) (Figure 1b; $p < 0.001$). When examining BMI, underweight and obese categories showed higher PHQ-8 scores of 9.5 (3.25–15.0) and 8.0 (4.0–13.0), compared to those in the normal (5.0 [2.0–9.0]) and overweight (5.0 [2.0–9.0]) groups, with the underweight group also reporting the highest GAD-7 scores across BMI categories (7.5 [3.0–12.75]) (Figure 1c; $p < 0.001$). Employment status analysis revealed lower depression and anxiety scores in retired (PHQ-8: 3.0 [1.0-7.0]; GAD-7: 1.0 [0.0-5.0]) and employed participants (PHQ-8: 6.0 [3.0-11.0]; GAD-7: 4.0 [1.0-8.0]), compared to higher scores in unemployed participants (PHQ-8: 10.0 [5.0-16.25]; GAD-7: 7.0 [3.0-13.0]) and students (PHQ-8: 10.5 [6.0-15.0]; GAD-7: 8.0 [4.0-12.0]) (Figure 1d; $p < 0.001$). Additionally, several other baseline variables such as marital status, living situation, ethnicity, and comorbidities were significantly correlated with PHQ-8 and GAD-7 scores, though some associations showed modest effect sizes, and others involved categories with limited sample sizes. Detailed results of all baseline variables are shown in

Supplementary Table 2.

Our Spearman correlation analysis highlighted a strong positive relationship between PHQ-8 and GAD-7 scores, with a Spearman correlation coefficient (ρ) of 0.79. Mood variables Valence and Arousal scores negatively correlated with both PHQ-8 (Valence: ρ=-0.51, Arousal: ρ=-0.49) and GAD-7 (Valence: ρ=-0.46, Arousal: ρ=-0.40). The PHQ-8 score exhibited negative correlations with variables related to steps (e.g., daily step counts [ρ=-0.24]) and activity (e.g., high activity duration [ρ=-0.22]) while showing positive correlations with variables rated to heart rates (e.g., nighttime heart rate [ρ=0.24]), and variability in sleep (e.g., sleep duration variability [ρ=0.23]). The GAD-7 score displayed similar but generally weaker correlations with Fitbit features compared to PHQ-8. Notably, the top three Fitbit features correlated with GAD-7 were all related to heart rate. Figure 2 respectively presents the top 20 variables correlated with PHQ-8 and GAD-7 and details of all correlations are shown in Supplementary Tables 3 and 4.

### 3.3 Clustering patterns associated with depression and anxiety severity

After dimensionality reduction through PCA (Figure 3a), K-means clustering identified three distinct clusters of Fitbit features (Figure 3b), primarily distinguished by the values of the first two PCs. PC1 is positively correlated with activity levels, while PC2 is correlated with heart rate levels, as indicated by the loadings of the original features (Figure 3a). Cluster 1 is characterized by participants with low activity and high heart rates, Cluster 2 by those with low activity and low heart rates, and Cluster 3 by high activity levels. Significant differences in PHQ-8 and GAD-7 scores were observed across these clusters (Figure 3c, $p<0.001$). Specifically, Cluster 1 participants reported the highest median scores (PHQ-8: 8.0 [4.0-13.0] and GAD-7: 6 [3.0-11.0]), followed by Cluster 2 (PHQ-8: 5.0 [2.0-10.0] and GAD-7: 4.0 [1.0-8.0]), with Cluster 3 presenting the lowest scores (PHQ-8: 4.0 [2.0-8.0] and GAD-7: 3.0 [1.0-7.0]).

### 3.4 Model performance for predicting depression and anxiety severity

The performance of XGBoost prediction models with different feature sets on the test data is displayed in Figures 4a and 4b. For PHQ-8 predictions, the model incorporating all features performed best ($R^2$=0.41, MAE=3.42); using only Fitbit, baseline, or mood features resulted in $R^2$ values of 0.15, 0.15, and 0.31 and MAEs of 4.19, 4.17, and 3.70, respectively; combinations of Fitbit and baseline features, Fitbit and mood features, and baseline and mood features yielded $R^2$ values of 0.22, 0.37, and 0.38 and MAEs of 3.98, 3.53, and 3.51, respectively. For GAD-7 predictions, the model with all features also had the highest performance ($R^2$=0.31, MAE=3.50); using only Fitbit, baseline, or mood features led to $R^2$ values of 0.08, 0.12, and 0.24 and MAEs of 4.13, 4.00, and 3.70, respectively; combinations of Fitbit and baseline features, Fitbit and mood features, and baseline and mood features achieved $R^2$ values of 0.14, 0.27, and 0.31 and MAEs of 3.93, 3.63, and 3.50, respectively.

Figures 4c and 4d visualize the impact of the top important features on the PHQ-8 and GAD-7 predictions. Key features in both models include Valence, Arousal, age, BMI, timing and variability of sleep, step cadence, and heart rate during the evening. The number of days with long sleep duration (potential hypertension), step count during the whole day and morning, and variability in step counts significantly influence the PHQ-8 predictions, whereas more heart rate-related features impacted GAD-7 predictions.

Notably, SHAP values illustrate a nonlinear relationship with daily steps for PHQ-8 predictions, where both extremely high and relatively low step counts are associated with higher PHQ-8 scores.

## 4. Discussion

This study explored indicators for depression and anxiety through digital phenotyping, analysing real-world gathered wearable and questionnaire data of 10,129 participants from a UK-based general population. Significant relationships were identified between depression and anxiety scores and various factors including mood, demographics, health metrics, and wearable-derived behavioural and physiological features. Machine learning models leveraging these variables can explain approximately 41% of the variance in depression scores and 31% in anxiety scores. These results highlight the potential of these variables as indicators for depression and anxiety and demonstrate the feasibility of using digital phenotyping to manage mental disorders in large, real-world populations.

The mood variables, Valence and Arousal, are negatively correlated with the severity of depression and anxiety, aligning with prior research that links lower levels of wakefulness and happiness to more severe symptoms (Moshe et al., 2021). Prediction models that incorporate only these two variables can explain 31% of the variance in PHQ-8 scores and 24% in GAD-7 scores (Figure 4a), highlighting the substantial impact of mood on predictions of depression and anxiety. These variables can be easily collected through simple smartphone-conducted mood tasks, such as using sliding buttons.

Both association analysis and SHAP values from prediction models demonstrated that age, gender, and BMI can be reliable indicators of depression and anxiety. Younger and female participants reported higher severity, consistent with research indicating a higher prevalence of depression and anxiety among younger individuals and women (Akhtar-Danesh and Landeen, 2007; Leach et al., 2008). Moreover, underweight and obese participants showed increased severity of depression, supporting epidemiological findings of a U-shaped relationship between depression and BMI (De Wit et al., 2009). This correlation may be influenced by biological factors, psychological stress, and lifestyle factors such as eating disorders and activity levels (De Wit et al., 2009). Inclusion of demographic and health variables in prediction models also significantly improved their overall performance, underscoring the importance of collecting this information of participants in the enrolment of future mHealth research.

Behavioural and physiological features extracted from commercial wearable devices also show significant correlations with depression and anxiety. Specifically, greater variability in sleep patterns and later sleep timings (i.e., onset and offset) are associated with higher depression severity, consistent with findings from previous mobile health studies (Fang et al., 2021; Robillard et al., 2015; Zhang et al., 2021a). For activity and step features, higher depression severity correlates with reduced activity duration, fewer steps, and slower step cadence, supporting past mHealth research (De Angel et al., 2022; McKercher et al., 2009; Zhang et al., 2022b). Increased activity, caloric consumption and steps during nighttime, and a delayed peak time of steps, indicating poor sleep and disrupted circadian rhythms, were associated with higher depression severity, as reported by previous studies (Smagula et al., 2018; Zhang et al., 2024b). Regarding heart rate features, depression severity is positively correlated with night-time heart rate and daily minimum heart rate, and negatively correlated with the range and variation of daily heart rate, consistent with previous studies (Schwerdtfeger and Friedrich-Mai, 2009; Siddi et al., 2023). Elevated nighttime and minimal daily heart rates are indicators of poor sleep (Alvaro et al., 2013), while the increased range and variation of daily heart rate can reflect more physical activities and better sleep (Alvaro et al., 2013; Kandola et al., 2019). For anxiety, similar associations have been observed, potentially due to the frequent comorbidity of depression and anxiety ($\rho=0.79$; Figure 2) (Tiller, 2012). Future randomized controlled trials are needed to explore unique indicators for anxiety and depression separately. Past clinical studies have closely linked anxiety disorders with heart rate variability (Chalmers et al., 2014), and our findings corroborate this, with many heart rate-related features ranking highly in the Spearman coefficients (Figure 2b) and the importance for predicting GAD-7 scores (Figure 4d). In addition, clustering analysis indicated the complex interplay of behavioural and physiological factors influencing depression and anxiety severity. Specifically, lower activity levels combined with higher heart rates are associated with more severe symptoms of depression and anxiety, while higher activity levels correspond to lower severity. These findings highlight the importance of integrating multimodal features to uncover nuanced relationships that may not be immediately apparent from unimodal analyses (e.g., not all low-activity patterns predict higher severity). SHAP values from the prediction model (Figure 4c) also illustrate the nonlinear relationships between some behavioural features and mental disorders, such as daily steps and sleep. This helps explain inconsistencies found in previous smaller studies (De Angel et al., 2022; Rohani et al., 2018) and highlights the necessity of using nonlinear models to capture complex relationships.

The results of the prediction models indicated that the best predictions for depression and anxiety severity were obtained by integrating mood, demographics and health data, along with behavioural and physiological features collected from wearable devices. When relying solely on Fitbit features, the models accounted for 15% and 8% of the variance in PHQ-8 and GAD-7 scores, respectively. This indicates that while certain Fitbit features are significantly associated with these disorders, their predictive effectiveness is limited. Previous mobile health studies have also reported limited predictive effectiveness when relying solely on passive features in cross-sectional predictions (Langholm et al., 2023; Pratap et al., 2019). This limitation is partly due to the influence of non-psychological factors such as work schedules, illness, and other life events (e.g., holidays) on individuals' behaviours and physiology, which can introduce bias into the predictions. Furthermore, the diverse manifestations of mental disorders (e.g., both insomnia and hypersomnia are manifestations of depression) and individual differences (e.g., average heart rate varies across individuals) can impact the performance of prediction models, indicating the need of individual modelling. Our ablation analysis underscored the improved performance of models when incorporating a variety of data types, highlighting the necessity of a multimodal approach for effective mental health prediction. Future research could consider adding other modalities associated with mental health, such as speech (Zhang et al., 2024a) and passive phone data (e.g., GPS (Zhang et al., 2022a), Bluetooth (Zhang et al., 2021b), and accelerometers (Zhang et al., 2022b)). However, it is also crucial to adapt the number of acceptable modalities based on varying degrees of user burden, to ensure maximum prediction accuracy and participant engagement in future studies.

This study has several limitations. Firstly, the data were collected during the COVID-19 pandemic, which may have introduced biases (Stewart et al., 2024; Sun et al., 2020) despite including some pandemic-related covariates; thus, our results need to be validated in post-pandemic datasets. Secondly, only a subset of demographic and health information (age, gender, and BMI) was initially collected at enrolment, with other variables gathered via an extended questionnaire. Due to participant attrition and engagement issues, we lost this information for some participants. Thirdly, to maintain interpretability, we used aggregated biweekly features in our machine learning models, potentially losing some day-to-day information. Future research will explore the use of daily features or raw data through sequential deep learning techniques to enhance prediction performance.

## 5. Conclusion

In conclusion, this study identified significant associations between depression and anxiety with mood, demographics, health metrics, and wearable-derived features in a large-scale UK-based general population. It also demonstrated the feasibility of using digital phenotyping and machine learning models to predict the self-reported severity of these disorders. Our findings highlight the benefit of integrating multiple data sources to enhance model performance, underscoring the importance of a multimodal approach for predicting mental health status. Despite some limitations, this research provides a foundational basis for future clinical applications using digital phenotyping approaches.

**Code availability**

The code used for the data processing and analysis can be made available upon reasonable requests.

**Data availability**

De-identified participant data are available for academic research purposes upon request to the corresponding author and the signing of a data access agreement.

**CRediT authorship contribution statement**
**Yuezhou Zhang:** Writing – original draft, Methodology, Formal analysis, Conceptualization.
**Callum Stewart:** Writing – review & editing, Software, Data curation. **Yatharth Ranjan:** Writing – review & editing, Software, Data curation. **Pauline Conde:** Writing – review & editing, Software. **Heet Sankesara:** Writing – review & editing. **Zulqarnain Rashid:** Writing – review & editing. **Shaoxiong Sun:** Writing – review & editing, Methodology. **Richard J.B. Dobson:** Writing – review & editing, Methodology, Project administration, Data curation, Conceptualization, Funding acquisition. **Amos A. Folarin:** Writing – review & editing, Methodology, Project administration, Data curation, Conceptualization, Funding acquisition.

**Declaration of competing interest**
Amos A. Folarin reports holding shares in Google, the parent company of Fitbit, which produces the wearable devices utilized in the Covid-Collab study to collect data. Fitbit advertised the Covid-Collab study in the UK Fitbit app. Neither Google nor Fitbit provided funding or devices for this study. All other authors declare no competing interests.

**Acknowledgements**
This study represents independent research partly funded by the National Institute for Health and Care Research (NIHR) Maudsley Biomedical Research Centre (IS-BRC-1215-20018 and NIHR203318) at South London and Maudsley NHS Foundation Trust, Medical Research Council, UK Research and Innovation, and King's College London. The views expressed in this paper are those of the authors and not necessarily those of the NIHR or the UK Department of Health and Social Care. Richard J.B. Dobson is supported by the following: (1) National Institute for Health and Care Research (NIHR) Biomedical Research Centre (BRC) at South London and Maudsley National Health Service (NHS) Foundation Trust and King's College London; (2) Health Data Research UK, which is funded by the UK Medical Research Council (MRC), Engineering and Physical Sciences Research Council, Economic and Social Research Council, Department of Health and Social Care (England), Chief Scientist Office of the Scottish Government Health and Social Care Directorates, Health and Social Care Research and Development Division (Welsh Government), Public Health Agency (Northern Ireland), British Heart Foundation, and Wellcome Trust; (3) the BigData@Heart Consortium, funded by the Innovative Medicines Initiative 2 Joint Undertaking (which receives support from the EU's Horizon 2020 research and innovation programme and European Federation of Pharmaceutical Industries and Associations [EFPIA], partnering with 20 academic and industry partners and European Society of Cardiology); (4) the NIHR University College London Hospitals BRC; (5) the NIHR BRC at South London and Maudsley (related to attendance at the American Medical Informatics Association) NHS Foundation Trust and King's College London; (6) the UK Research and Innovation (UKRI) London Medical Imaging & Artificial Intelligence Centre for Value Based Healthcare (AI4VBH); and (7) the NIHR Applied Research Collaboration (ARC) South London at King's College Hospital NHS Foundation Trust.

# Reference

Abd-Alrazaq, A., AlSaad, R., Aziz, S., Ahmed, A., Denecke, K., Househ, M., Farooq, F., Sheikh, J., 2023. Wearable artificial intelligence for anxiety and depression: scoping review. Journal of Medical Internet Research. 25, e42672.

Akhtar-Danesh, N., Landeen, J., 2007. Relation between depression and sociodemographic factors. International journal of mental health systems. 1, 1-9.

Alvaro, P.K., Roberts, R.M., Harris, J.K., 2013. A systematic review assessing bidirectionality between sleep disturbances, anxiety, and depression. Sleep. 36, 1059-1068.

Anikwe, C.V., Nweke, H.F., Ikegwu, A.C., Egwuonwu, C.A., Onu, F.U., Alo, U.R., Teh, Y.W., 2022. Mobile and wearable sensors for data-driven health monitoring system: State-of-the-art and future prospect. Expert Systems with Applications. 202, 117362.

Bayram, N., Bilgel, N., 2008. The prevalence and socio-demographic correlations of depression, anxiety and stress among a group of university students. Social psychiatry and psychiatric epidemiology. 43, 667-672.

Ben-Zeev, D., Young, M.A., 2010. Accuracy of hospitalized depressed patients' and healthy controls' retrospective symptom reports: an experience sampling study. The Journal of nervous and mental disease. 198, 280-285.

Benjamini, Y., Hochberg, Y., 1995. Controlling the false discovery rate: a practical and powerful approach to multiple testing. Journal of the Royal statistical society: series B (Methodological). 57, 289-300.

Bufano, P., Laurino, M., Said, S., Tognetti, A., Menicucci, D., 2023. Digital Phenotyping for Monitoring Mental Disorders: Systematic Review. Journal of Medical Internet Research. 25, e46778.

Chalmers, J.A., Quintana, D.S., Abbott, M.J.-A., Kemp, A.H., 2014. Anxiety disorders are associated with reduced heart rate variability: a meta-analysis. Frontiers in psychiatry. 5, 80.

Chen, T., Guestrin, C., 2016. Xgboost: A scalable tree boosting system, Proceedings of the 22nd acm sigkdd international conference on knowledge discovery and data mining, pp. 785-794.

Cimpean, D., Drake, R., 2011. Treating co-morbid chronic medical conditions and anxiety/depression. Epidemiology and psychiatric sciences. 20, 141-150.

Collaborators, G.M.D., 2022. Global, regional, and national burden of 12 mental disorders in 204 countries and territories, 1990–2019: a systematic analysis for the Global Burden of Disease Study 2019. The Lancet Psychiatry. 9, 137-150.

Currey, D., Torous, J., 2022. Digital phenotyping correlations in larger mental health samples: analysis and replication. BJPsych Open. 8, e106.

De Angel, V., Lewis, S., White, K., Oetzmann, C., Leightley, D., Oprea, E., Lavelle, G., Matcham, F., Pace, A., Mohr, D.C., 2022. Digital health tools for the passive monitoring of depression: a systematic review of methods. NPJ digital medicine. 5, 3.

De Wit, L.M., Van Straten, A., Van Herten, M., Penninx, B.W., Cuijpers, P., 2009. Depression and body mass index, a u-shaped association. BMC public health. 9, 1-6.

Fang, Y., Forger, D.B., Frank, E., Sen, S., Goldstein, C., 2021. Day-to-day variability in sleep parameters and depression risk: a prospective cohort study of training physicians. NPJ digital medicine. 4, 28.

Granato, D., Santos, J.S., Escher, G.B., Ferreira, B.L., Maggio, R.M., 2018. Use of principal component analysis (PCA) and hierarchical cluster analysis (HCA) for multivariate association between bioactive compounds and functional properties in foods: A critical perspective. Trends in Food Science & Technology. 72, 83-90.

Huckins, J.F., DaSilva, A.W., Wang, W., Hedlund, E., Rogers, C., Nepal, S.K., Wu, J., Obuchi, M., Murphy, E.I., Meyer, M.L., 2020. Mental health and behavior of college students during the early phases of the COVID-19 pandemic: Longitudinal smartphone and ecological momentary assessment study. Journal of medical Internet research. 22, e20185.

Insel, T.R., 2017. Digital phenotyping: technology for a new science of behavior. Jama. 318, 1215-1216.

Kandola, A., Ashdown-Franks, G., Hendrikse, J., Sabiston, C.M., Stubbs, B., 2019. Physical activity and depression: Towards understanding the antidepressant mechanisms of physical activity. Neuroscience & Biobehavioral Reviews. 107, 525-539.

Kroenke, K., Strine, T.W., Spitzer, R.L., Williams, J.B., Berry, J.T., Mokdad, A.H., 2009. The PHQ-8 as a measure of current depression in the general population. Journal of affective disorders. 114, 163-173.

Langholm, C., Breitinger, S., Gray, L., Goes, F., Walker, A., Xiong, A., Stopel, C., Zandi, P., Frye, M.A., Torous, J., 2023. Classifying and clustering mood disorder patients using smartphone data from a feasibility study. npj Digital Medicine. 6, 238.

Leach, L.S., Christensen, H., Mackinnon, A.J., Windsor, T.D., Butterworth, P., 2008. Gender differences in depression and anxiety across the adult lifespan: the role of psychosocial mediators. Social psychiatry and psychiatric epidemiology. 43, 983-998.

Liu, Q., He, H., Yang, J., Feng, X., Zhao, F., Lyu, J., 2020. Changes in the global burden of depression from 1990 to 2017: Findings from the Global Burden of Disease study. Journal of psychiatric research. 126, 134-140.

Lundberg, S.M., Lee, S.-I., 2017. A unified approach to interpreting model predictions. Advances in neural information processing systems. 30.

McKercher, C.M., Schmidt, M.D., Sanderson, K.A., Patton, G.C., Dwyer, T., Venn, A.J., 2009. Physical activity and depression in young adults. American journal of preventive medicine. 36, 161-164.

Mohr, D.C., Shilton, K., Hotopf, M., 2020. Digital phenotyping, behavioral sensing, or personal sensing: names and transparency in the digital age. NPJ digital medicine. 3, 45.

Mohr, D.C., Zhang, M., Schueller, S.M., 2017. Personal sensing: understanding mental health using ubiquitous sensors and machine learning. Annual review of clinical psychology. 13, 23-47.

Moshe, I., Terhorst, Y., Opoku Asare, K., Sander, L.B., Ferreira, D., Baumeister, H., Mohr, D.C., Pulkki-Råback, L., 2021. Predicting symptoms of depression and anxiety using smartphone and wearable data. Frontiers in psychiatry. 12, 625247.

Murray, C.J., Vos, T., Lozano, R., Naghavi, M., Flaxman, A.D., Michaud, C., Ezzati, M., Shibuya, K., Salomon, J.A., Abdalla, S., 2012. Disability-adjusted life years (DALYs) for 291 diseases and injuries in 21 regions, 1990–2010: a systematic analysis for the Global Burden of Disease Study 2010. The lancet. 380, 2197-2223.

Okobi, O.E., Sobayo, T.O., Arisoyin, A.E., Adeyemo, D.A., Olaleye, K.T., Nelson, C.O., Sanusi, I.A., Salawu, M.A., Akinsete, A.O., Emore, E., 2023. Association between the use of wearable devices and physical activity among US adults with depression and anxiety: Evidence from the 2019 and 2020 Health Information National Trends Survey. Cureus. 15.

Oladeji, B.D., Gureje, O., 2016. Brain drain: a challenge to global mental health. BJPsych international. 13, 61-63.

Ostertagova, E., Ostertag, O., Kováč, J., 2014. Methodology and application of the Kruskal-Wallis test.

Applied mechanics and materials. 611, 115-120.

Pratap, A., Atkins, D.C., Renn, B.N., Tanana, M.J., Mooney, S.D., Anguera, J.A., Areán, P.A., 2019. The accuracy of passive phone sensors in predicting daily mood. Depression and anxiety. 36, 72-81.

Ranjan, Y., Rashid, Z., Stewart, C., Conde, P., Begale, M., Verbeeck, D., Boettcher, S., Dobson, R., Folarin, A., Consortium, R.-C., 2019. RADAR-base: open source mobile health platform for collecting, monitoring, and analyzing data using sensors, wearables, and mobile devices. JMIR mHealth and uHealth. 7, e11734.

Robillard, R., Hermens, D.F., Naismith, S.L., White, D., Rogers, N.L., Ip, T.K., Mullin, S.J., Alvares, G.A., Guastella, A.J., Smith, K.L., 2015. Ambulatory sleep-wake patterns and variability in young people with emerging mental disorders. Journal of Psychiatry and Neuroscience. 40, 28-37.

Rohani, D.A., Faurholt-Jepsen, M., Kessing, L.V., Bardram, J.E., 2018. Correlations between objective behavioral features collected from mobile and wearable devices and depressive mood symptoms in patients with affective disorders: systematic review. JMIR mHealth and uHealth. 6, e9691.

Santomauro, D.F., Herrera, A.M.M., Shadid, J., Zheng, P., Ashbaugh, C., Pigott, D.M., Abbafati, C., Adolph, C., Amlag, J.O., Aravkin, A.Y., 2021. Global prevalence and burden of depressive and anxiety disorders in 204 countries and territories in 2020 due to the COVID-19 pandemic. The Lancet. 398, 1700-1712.

Schwerdtfeger, A., Friedrich-Mai, P., 2009. Social interaction moderates the relationship between depressive mood and heart rate variability: evidence from an ambulatory monitoring study. Health Psychology. 28, 501.

Siddi, S., Bailon, R., Gine-Vazquez, I., Matcham, F., Lamers, F., Kontaxis, S., Laporta, E., Garcia, E., Lombardini, F., Annas, P., 2023. The usability of daytime and night-time heart rate dynamics as digital biomarkers of depression severity. Psychological medicine. 53, 3249-3260.

Smagula, S.F., Krafty, R.T., Thayer, J.F., Buysse, D.J., Hall, M.H., 2018. Rest-activity rhythm profiles associated with manic-hypomanic and depressive symptoms. Journal of psychiatric research. 102, 238-244.

Spearman, C., 1961. The proof and measurement of association between two things.

Spitzer, R.L., Kroenke, K., Williams, J.B., Löwe, B., 2006. A brief measure for assessing generalized anxiety disorder: the GAD-7. Archives of internal medicine. 166, 1092-1097.

Stewart, C., Ranjan, Y., Conde, P., Rashid, Z., Sankesara, H., Bai, X., Dobson, R.J., Folarin, A.A., 2021. Investigating the use of digital health technology to monitor COVID-19 and its effects: protocol for an observational study (COVID Collab Study). JMIR research protocols. 10, e32587.

Stewart, C., Ranjan, Y., Conde, P., Sun, S., Zhang, Y., Rashid, Z., Sankesara, H., Cummins, N., Laiou, P., Bai, X., 2024. Physiological presentation and risk factors of long COVID in the UK using smartphones and wearable devices: a longitudinal, citizen science, case–control study. The Lancet Digital Health.

Sun, S., Folarin, A.A., Ranjan, Y., Rashid, Z., Conde, P., Stewart, C., Cummins, N., Matcham, F., Dalla Costa, G., Simblett, S., 2020. Using smartphones and wearable devices to monitor behavioral changes during COVID-19. Journal of medical Internet research. 22, e19992.

Sun, S., Folarin, A.A., Zhang, Y., Cummins, N., Garcia-Dias, R., Stewart, C., Ranjan, Y., Rashid, Z., Conde, P., Laiou, P., 2023. Challenges in Using mHealth Data From Smartphones and Wearable Devices to Predict Depression Symptom Severity: Retrospective Analysis. Journal of medical Internet research. 25, e45233.

Syakur, M., Khotimah, B.K., Rochman, E., Satoto, B.D., 2018. Integration k-means clustering method and elbow method for identification of the best customer profile cluster, IOP conference series: materials science and engineering. IOP Publishing, p. 012017.

Tiller, J.W., 2012. Depression and anxiety. Medical Journal of Australia. 1.

Torous, J., Kiang, M.V., Lorme, J., Onnela, J.-P., 2016. New tools for new research in psychiatry: a scalable and customizable platform to empower data driven smartphone research. JMIR mental health. 3, e5165.

Torous, J., Myrick, K.J., Rauseo-Ricupero, N., Firth, J., 2020. Digital mental health and COVID-19: using technology today to accelerate the curve on access and quality tomorrow. JMIR mental health. 7, e18848.

Ueafuea, K., Boonnag, C., Sudhawiyangkul, T., Leelaarporn, P., Gulistan, A., Chen, W., Mukhopadhyay, S.C., Wilaiprasitporn, T., Piyayotai, S., 2020. Potential applications of mobile and wearable devices for psychological support during the COVID-19 pandemic: a review. IEEE Sensors Journal. 21, 7162-7178.

van Rijsbergen, G.D., Bockting, C.L., Burger, H., Spinhoven, P., Koeter, M.W., Ruhé, H.G., Hollon, S.D., Schene, A.H., 2013. Mood reactivity rather than cognitive reactivity is predictive of depressive relapse: a randomized study with 5.5-year follow-up. Journal of Consulting and Clinical Psychology. 81, 508.

Wang, R., Chen, F., Chen, Z., Li, T., Harari, G., Tignor, S., Zhou, X., Ben-Zeev, D., Campbell, A.T., 2014. StudentLife: assessing mental health, academic performance and behavioral trends of college students using smartphones, Proceedings of the 2014 ACM international joint conference on pervasive and ubiquitous computing, pp. 3-14.

Xu, X., Liu, X., Zhang, H., Wang, W., Nepal, S., Sefidgar, Y., Seo, W., Kuehn, K.S., Huckins, J.F., Morris, M.E., 2023. GLOBEM: cross-dataset generalization of longitudinal human behavior modeling. Proceedings of the ACM on Interactive, Mobile, Wearable and Ubiquitous Technologies. 6, 1-34.

Zhang, Y., Folarin, A.A., Dineley, J., Conde, P., de Angel, V., Sun, S., Ranjan, Y., Rashid, Z., Stewart, C., Laiou, P., 2024a. Identifying depression-related topics in smartphone-collected free-response speech recordings using an automatic speech recognition system and a deep learning topic model. Journal of affective disorders. 355, 40-49.

Zhang, Y., Folarin, A.A., Sun, S., Cummins, N., Bendayan, R., Ranjan, Y., Rashid, Z., Conde, P., Stewart, C., Laiou, P., 2021a. Relationship between major depression symptom severity and sleep collected using a wristband wearable device: multicenter longitudinal observational study. JMIR mHealth and uHealth. 9, e24604.

Zhang, Y., Folarin, A.A., Sun, S., Cummins, N., Ranjan, Y., Rashid, Z., Conde, P., Stewart, C., Laiou, P., Matcham, F., 2021b. Predicting depressive symptom severity through individuals' nearby bluetooth device count data collected by mobile phones: preliminary longitudinal study. JMIR mHealth and uHealth. 9, e29840.

Zhang, Y., Folarin, A.A., Sun, S., Cummins, N., Ranjan, Y., Rashid, Z., Stewart, C., Conde, P., Sankesara, H., Laiou, P., 2024b. Longitudinal Assessment of Seasonal Impacts and Depression Associations on Circadian Rhythm Using Multimodal Wearable Sensing: Retrospective Analysis. Journal of Medical Internet Research. 26, e55302.

Zhang, Y., Folarin, A.A., Sun, S., Cummins, N., Vairavan, S., Bendayan, R., Ranjan, Y., Rashid, Z., Conde, P., Stewart, C., 2022a. Longitudinal relationships between depressive symptom severity and phone-measured mobility: dynamic structural equation modeling study. JMIR mental health. 9, e34898.

Zhang, Y., Folarin, A.A., Sun, S., Cummins, N., Vairavan, S., Qian, L., Ranjan, Y., Rashid, Z., Conde, P., Stewart, C., 2022b. Associations between depression symptom severity and daily-life gait characteristics derived from long-term acceleration signals in real-world settings: retrospective analysis. JMIR mHealth and uHealth. 10, e40667.

Zhong, S., Yang, X., Pan, Z., Fan, Y., Chen, Y., Yu, X., Zhou, L., 2023. The usability, feasibility, acceptability, and efficacy of digital mental health services in the COVID-19 pandemic: scoping review, systematic review, and meta-analysis. JMIR Public Health and Surveillance. 9, e43730.

**Table 1.** A summary of characteristics of selected participants in this study.

| Characteristic | Selected Cohort |
|---|---|
| Number of Participants, n (%) | 10129 |
| PHQ-8 score, median (IQR) | 6.0 (3.0-10.0) |
| GAD-7 score, median (IQR) | 4.0 (1.0-8.0) |
| Age, median (IQR) | 48.0 (36.0-58.0) |
| Gender=Female | 7008 (69.3%) |
| BMI, median (IQR) | 26.2 (23.2-30.3) |
| **Extended Questionnaire (N = 4115)** | |
| Employment, n (%) | |
|    Employed | 2985 (72.5%) |
|    Retired | 755 (18.3%) |
|    Student | 170 (4.1%) |
|    Unemployed | 211 (5.1%) |
| Marital Status, n (%) | |
|    In a relationship | 3009 (73.1%) |
|    Single or separated | 1084 (26.3%) |
| Living Situation, n (%) | |
|    With family/partner/housemate | 3568 (86.7%) |
|    Alone | 544 (13.2%) |
| Ethnicity, n (%) | |
|    White | 3760 (91.4%) |
|    Asian | 81 (2.0%) |
|    Black | 24 (0.6%) |
|    Mixed | 45 (1.1%) |
|    Other | 5 (0.1%) |
|    Unknown | 215 (5.2%) |
| Comorbidities, n (%) | |
|    Asthma | 802 (19.4%) |
|    Cancer | 210 (5.1%) |
|    Diabetes (Type 2) | 140 (3.4%) |
|    Cardiovascular Diseases | 135 (3.3%) |
|    Hypertension | 530 (12.8%) |
|    Obesity | 335 (8.1%) |

Note that only a portion of the participants completed the extended questionnaires due to variations in engagement. Additionally, a small portion of responses to each characteristic could not be categorized due to user entry errors, resulting in percentages that do not sum to 100% for each characteristic.

**Figure Legends**

**Figure 1.** Distribution of depression (PHQ-8) and anxiety (GAD-7) scores across demographic and health characteristics: (a) age, (b) gender, (c) body mass index (BMI), and (d) employment status.

**Figure 2.** Top 20 correlations with depression (PHQ-8) and anxiety (GAD-7) scores: (a) correlations with PHQ-8, and (b) correlations with GAD-7.

**Figure 3.** Participant clusters of Fitbit patterns and corresponding boxplots of depression (PHQ-8) and anxiety (GAD-7) scores. (a) Loadings of original Fitbit features on principal components (PCs). (b) K-means clustering results are visualized on the first two PCs, which are related to the activity level and heart rate level, respectively. (c) Boxplots of PHQ-8 and GAD-7 scores across identified clusters.

**Figure 4.** Prediction performance of models with different feature sets for predicting depression (PHQ-8) and anxiety (GAD-7) scores, and visualizations of feature importance. (a) Coefficient of determination ($R^2$), (b) Mean absolute error (MAE) (c) SHAP plots for PHQ-8 prediction model with all features, d) SHAP plots for GAD-7 prediction models with all features.

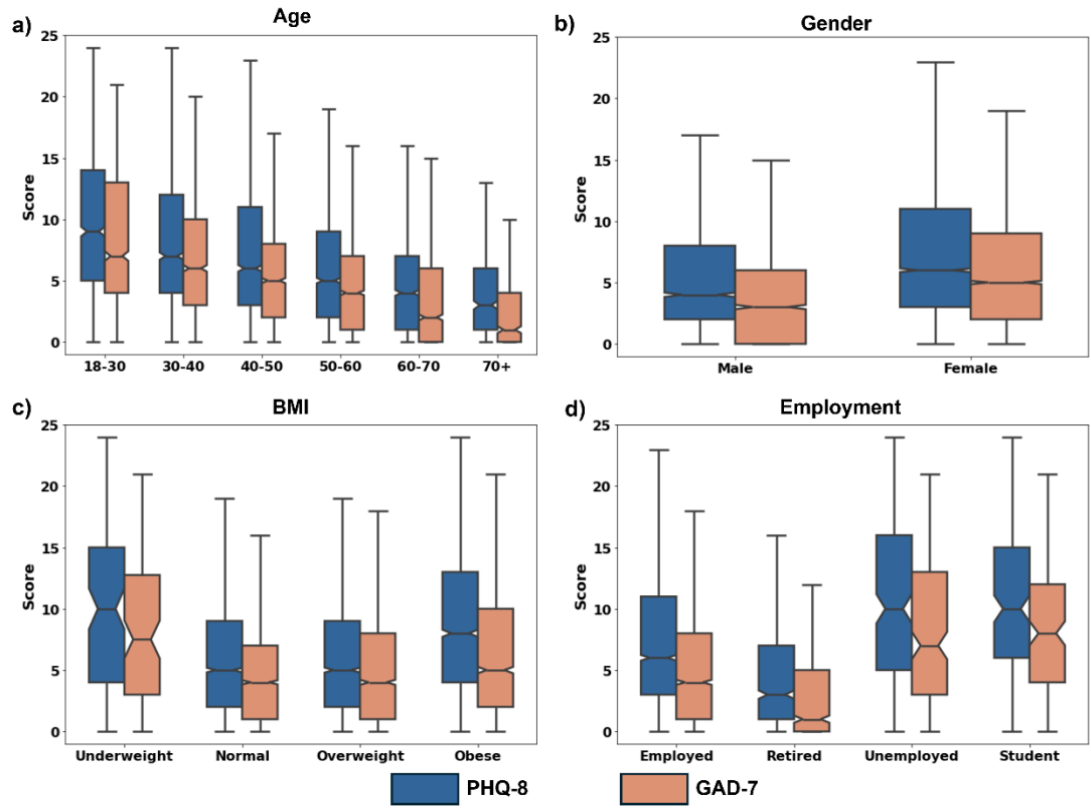

**Figure 1. Distribution of depression (PHQ-8) and anxiety (GAD-7) scores across demographic and health characteristics: (a) age, (b) gender, (c) body mass index (BMI), and (d) employment status.**

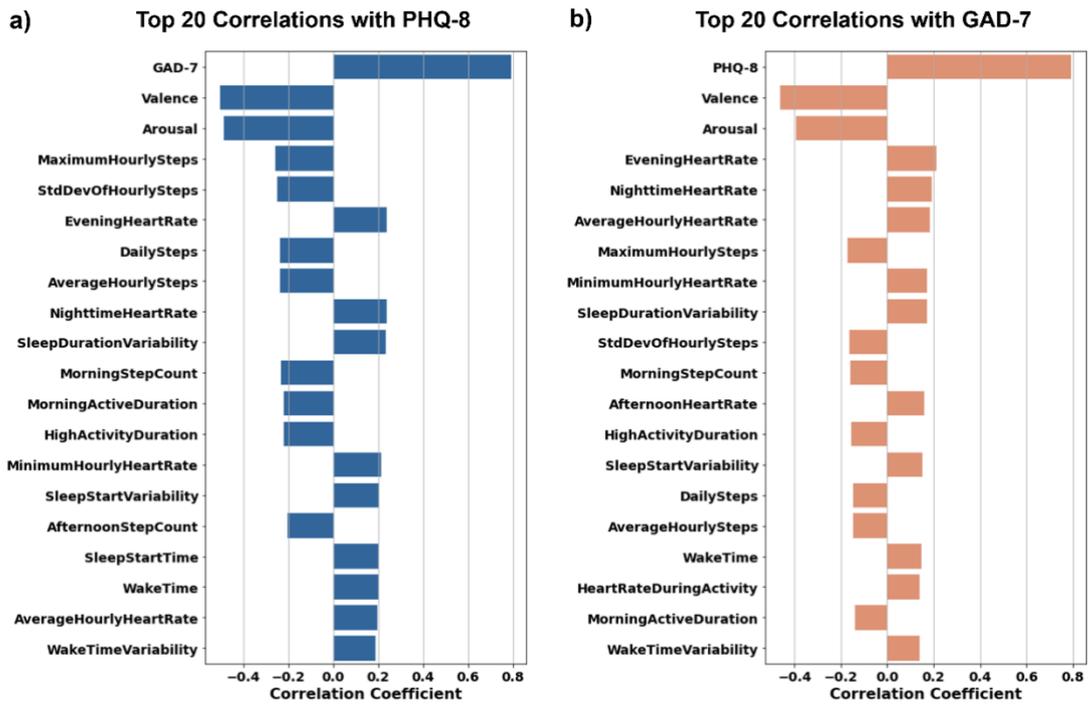

**Figure 2. Top 20 significant correlations with depression (PHQ-8) and anxiety (GAD-7) scores: (a) correlations with PHQ-8, and (b) correlations with GAD-7.**

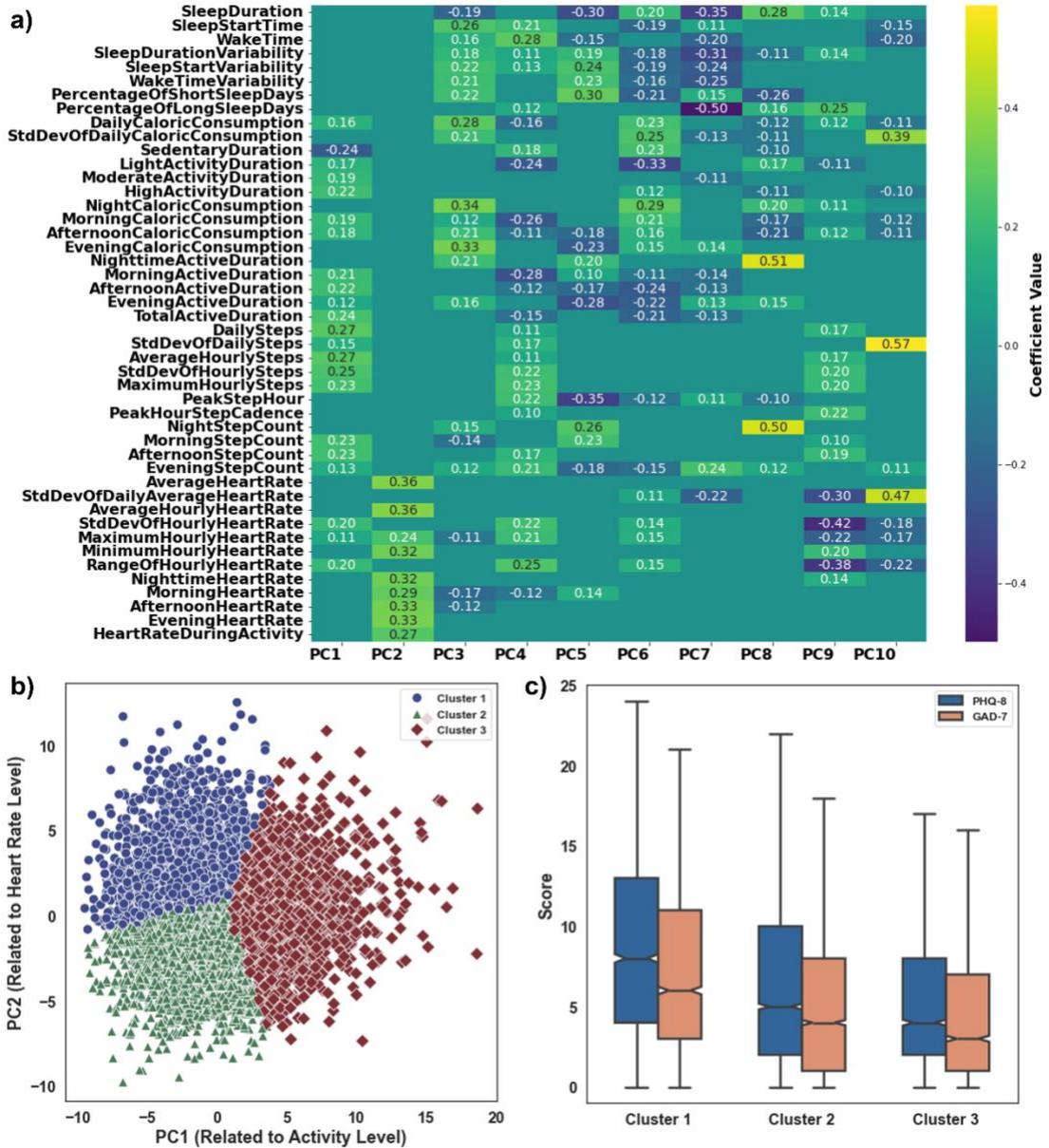

**Figure 3. Participant clusters of Fitbit patterns and corresponding boxplots of depression (PHQ-8) and anxiety (GAD-7) scores. (a) Loadings of original Fitbit features on principal components (PCs). (b) K-means clustering results are visualized on the first two PCs, which are related to the activity level and heart rate level, respectively. (c) Boxplots of PHQ-8 and GAD-7 scores across identified clusters.**

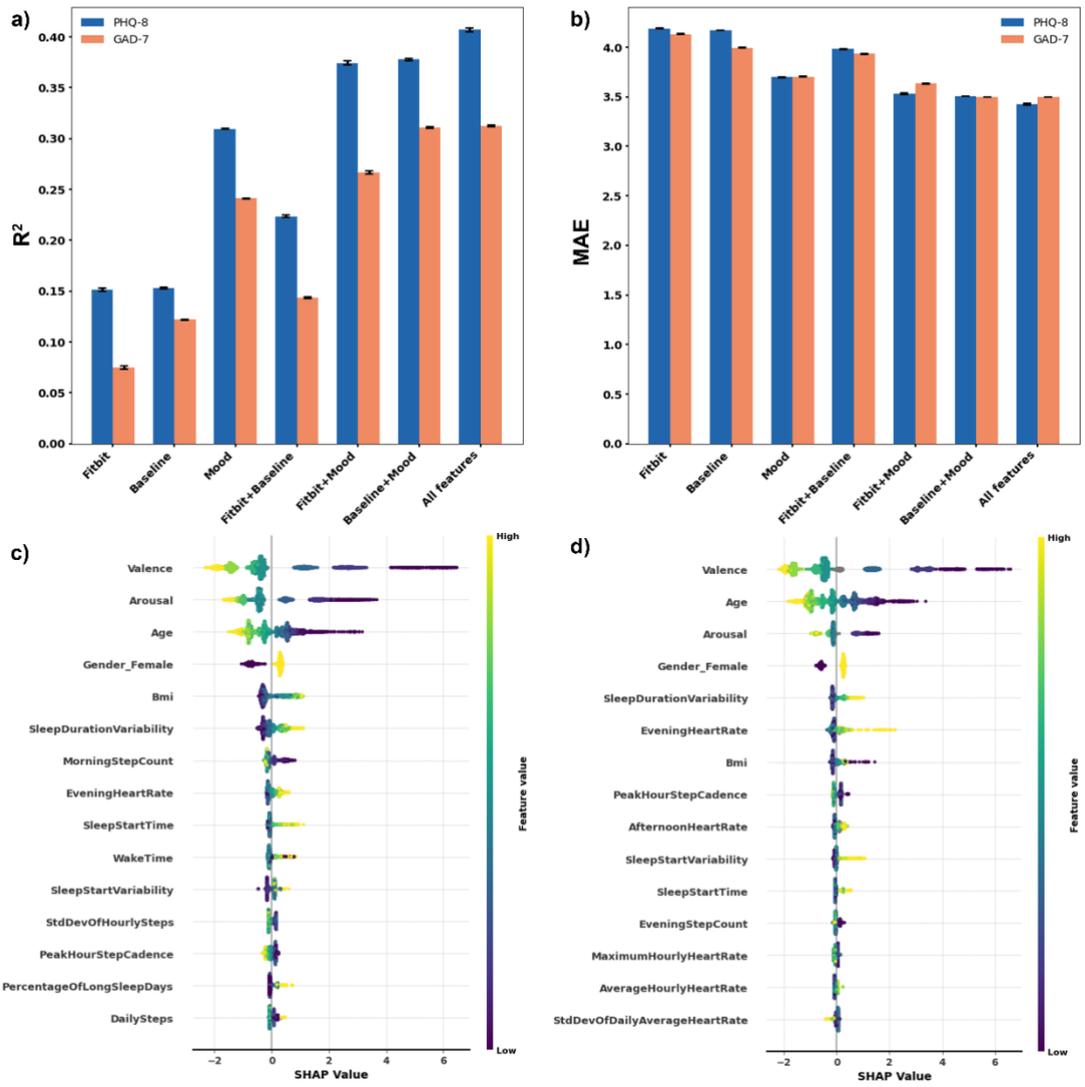

Figure 4. Prediction performance of models with different feature sets for predicting depression (PHQ-8) and anxiety (GAD-7) scores, and visualizations of feature importance. (a) Coefficient of determination ($R^2$), (b) Mean absolute error (MAE) (c) SHAP plots for PHQ-8 prediction model with all features, d) SHAP plots for GAD-7 prediction models with all features.